\documentclass[review,12pt]{elsarticle}

\usepackage{amssymb}

\usepackage{longtable}
\usepackage{bm}
\usepackage{multirow}
\usepackage{hyperref}
\usepackage{ulem}
\hypersetup{colorlinks,citecolor=blue,filecolor=blue,linkcolor=blue,urlcolor=blue}
\usepackage{color}
\usepackage{dcolumn}
\usepackage{tabularx}
\usepackage[usenames,dvipsnames]{xcolor}

\definecolor{awesome}{rgb}{1.0, 0.13, 0.32}
\newcommand{\iso}[2]{$^{#1}$#2}

\newcommand{\FSUDAQ}[0]{\textsf{FSUDAQ}}
\newcommand{\BoxScore}[0]{\textsf{BoxScore}}

\journal{}

\begin{document}

\begin{frontmatter}


\title{FSUDAQ - A general purpose GUI data acquisition program for the CAEN x725, x730, x740 digitizers}


\author[inst1]{T.~L.~Tang}
\ead{rtang@fsu.edu}
\address[inst1]{Department of Physics, Florida State University, Tallahassee, Florida 32306, USA}

\begin{abstract}
\FSUDAQ~is a versatile, multi-threaded, lightweight data acquisition software with a graphical user interface, designed to fully utilize the capabilities of first-generation CAEN x725, x730, and x740 series digitizers equipped with various Digital Pulse Processing (DPP) firmware, including Pulse-Height Analysis (PHA), Pulse-Shape Discrimination (PSD), and Charge-Digital Conversion (QDC). It emphasizes user-friendliness, stability, scalability, high throughput, and low latency. The software includes features such as an online waveform scope, scalar panel, and real-time single spectrum display for each input channel, along with an online event builder and analyzer capable of generating 1D and 2D histograms and applying graphical cuts. Users can also create and integrate custom online analyzers to meet specific experimental requirements. \FSUDAQ~has been successfully tested at the John D. Fox Laboratory at FSU, including with the Encore, ANASEN, and Super-Enge Split-Pole Spectrograph experiments. In terms of performance, \FSUDAQ~can handle up to approximately 500k triggers per second per channel without waveform recording, or data rates of around 65 MB/s per optical fiber, with or without waveform recording.

\end{abstract}



\begin{keyword}


\end{keyword}

\end{frontmatter}


\section{Introduction}

The CAEN x725, x730, and x740 series digitizers~\cite{CAENDigitizer} have gained widespread popularity as data acquisition (DAQ) devices in nuclear physics. 
With the development of \BoxScore~\cite{BoxScore}, the need for a versatile, native graphical user interface (GUI) DAQ program became evident. Such a program would provide a user-friendly interface while supporting real-time event building, data analysis, and serving as a multi-purpose DAQ system for these digitizers. Although CAEN offers \texttt{CoMPASS} for similar purposes, its broad support for various CAEN digitizers and firmware often sacrifices conveniences and features specific to nuclear physics applications.

Nuclear physics experiments often impose specialized requirements. One common need is the ability to record waveforms alongside energy and timestamp data to enable complex waveform analysis. Additionally, applying graphical gates to 2D histograms for real-time analysis is crucial for monitoring the status of ongoing experiments. Another requirement is full control over the digitizer, such as customizing the trigger scheme. Many of these settings are hidden or difficult to access in \texttt{CoMPASS}. While \texttt{CoMPASS} allows manipulation of registers through a ``free write" file, which is not intuitive for users.

\FSUDAQ~\cite{FSUDAQ} addresses these challenges by offering a range of features designed to enhance user control and efficiency. Its core functions include: 1) the ability to search for and open connected digitizers, even in cases where the Phase-Locked Loop fails, a scenario in which \texttt{CoMPASS} cannot establish a connection; 2) comprehensive register control via the GUI; 3) a synchronization helper to ensure consistent timing across multiple digitizers; 4) flexible data acquisition options, allowing users to capture data with or without waveform recording; 5) real-time scalar information retrieval through rapid buffer scanning, enabling instantaneous monitoring of all channels; and 6) a waveform scope that assists in fine-tuning Digital Pulse Processing (DPP) parameters for optimal signal processing.

In addition to its core functionalities, \FSUDAQ~offers several supplementary features that enhance its versatility and performance. These include: A) the ability to output information, such as scalar values, to an \textsf{InfluxDB} database~\cite{influx} via network connection; B) automatic \textsf{elog}~\cite{elog} entry creation, streamlining the documentation of experimental details and improving reproducibility and record-keeping; C) a real-time event builder and analyzer that can handle data from multiple digitizers simultaneously, allowing for concurrent data acquisition and online analysis in complex experimental setups; and D) circular memory to efficiently limit the usage of the hosting computer's memory.

The design of the DAQ system is driven by several key objectives: I) Providing a user-friendly experience with intuitive control, easy installation, and straightforward modification and maintenance of the code; II) Achieving high throughput, with a capability of approximately 65~MB/s per optical cable; III) Ensuring low latency and high responsiveness; IV) Scaling to accommodate an arbitrary number of digitizers; V) Maintaining stability and crash-free operation; and VI) Being future-proof. To meet these objectives, \FSUDAQ~is built using C++ and Qt6~\cite{Qt6}, which offer flexible memory management and a rich set of GUI elements, making them ideal for developing a robust DAQ system. Qt6 is also well-documented and supported by an active community, ensuring its relevance and usability in the foreseeable future.

This report provides a comprehensive overview of \FSUDAQ, starting with an outline of its architecture, focusing on its core components and the implementation of multi-threaded data reading. It then explores the customization options available for online analysis. Following this, the report presents several real-world applications, including its deployment at the Encore active-target~\cite{Encore}, the ANASEN~\cite{ANASEN}, and the Super-Enge Split-Pole Spectrograph~\cite{SPS} at FSU, as well as the ATLAS in-flight separator~\cite{RAISOR} at Argonne National Laboratory. These examples illustrate \FSUDAQ's versatility in different experimental environments. The report further evaluates \FSUDAQ's capabilities and performance, assessing its effectiveness in fulfilling the requirements of data acquisition and analysis. Finally, it explores the potential for integrating additional DPP firmware into \FSUDAQ, highlighting future expansion opportunities.

\section{Architecture of \FSUDAQ }

\FSUDAQ~is developed on Ubuntu 22.04 using C++ with Qt 6.2, taking advantage of Qt's robust and extensive GUI elements and multi-threading capabilities~\cite{Qt6}. Notably, \FSUDAQ~does not rely on the Qt Creator for GUI design and programming. Instead, it utilizes simple plain C++ classes derived from the \texttt{QWidget} class. This approach has better code readability, simplifies the development environment, and reduces code maintenance efforts. 

\FSUDAQ~has been tested with Ubuntu 22.04, 23.04, and 24.04 operating systems. Additionally, it has been verified to work with direct USB connections and optical fiber connection via the CAEN A3818 and A5818 optical fiber to PCIe adapters. Furthermore, \FSUDAQ~has been successfully installed and tested on Raspberry Pi 4B and 5 using the CAEN A4818 optical fiber to USB 3.0 adapter.

\begin{table}[h!]
\centering
\begin{tabular}{lll}
Library    & Version  \\
\hline
CAENVMELib    & 3.3+    \\
CAENComm      & 1.5.3+  \\
CAENDigitizer & 2.17.1+  \\
CAEN A3818 Driver$^{*}$ & 1.6.4+ \\
CAEN A4818 Driver$^{*}$ & 1.0+ \\
CAEN A5818 Driver$^{*}$ & 1.0.3+ \\
CAEN USB Driver$^{*}$ & 1.5.5+ \\
\texttt{qt6-base-dev} & 6.4.2+ \\
\texttt{libqt6charts6-dev} &  6.4.2+  \\
\texttt{libcurl4-openssl-dev} & 7.81.0+ \\
\end{tabular}
\caption{CAEN, Qt6, and CURL libraries needed for the \texttt{FSUDAQ}. The $^*$ libraries are optional depends on which interface is being used.}
\label{table:CAENLib}
\end{table}

The CAEN, Qt6, and CURL~\cite{curl4} libraries listed on Table.~\ref{table:CAENLib} are required to communicate with the digitizers, build GUI, and communicate \textsf{InfluxDB} via internet. Additional packages for \textsf{elog}~\cite{elog} is optional. The Qt6 and CURL packages are readily available in the official software repositories of Ubuntu 22.04+, and can be installed via the `apt' command. This makes \FSUDAQ~can be easily deployed on any Ubuntu 22.04 or later machine. Unlike \BoxScore, \FSUDAQ~does not rely on the CERN ROOT libraries~\cite{ROOT6} for histograms. Instead, histograms are created using the open-source \texttt{QCustomPlot} library~\cite{QCustomPlot}. By utilizing readily available packages and open-source libraries, \FSUDAQ~simplifies the setup process and ensures compatibility with a wide range of Ubuntu-based systems.

The backbone of \FSUDAQ~consists of four core C++ classes : \texttt{Reg} (short for register),  \texttt{Data}, \texttt{Digitizer}, and \texttt{MultiBuilder}. The four classes required only the CAEN libraries and are independent of Qt6, CURL, and \textsf{elog}. A standalone data acquisition system can be built using these classes. The GUI elements of \FSUDAQ~were built upon the Qt6 and CURL libraries.

\subsection{\textnormal{\texttt{Reg}} class}

The \texttt{Reg} class was designed to manage the digitizer registers. It effectively stores and organizes crucial register details, including addresses, read-write permissions, maximum values (if applicable), step sizes, and etc.. For instance, a unit of the DPP-PHA pre-trigger (register address 0x1038) corresponds to 8 samples, with each sample varying from 2 to 16 ns based on the digitizer model. This information is encapsulated within the \texttt{PreTrigger} object, enhancing usability and reducing the likelihood of coding errors. Furthermore, each DPP register is assigned to its respective DPP namespace to prevent conflicts between registers with identical names in different DPP firmwares.

All register values are stored as an unsigned 32-bit integer array of size 2048 within the \texttt{Digitizer} class (see section~\ref{sec::DigitizerClass}). To efficiently map between register addresses and their corresponding indices within the settings array, the \texttt{Reg::Index(address)} method calculates the index for a given address, while the \texttt{Reg::CalAddress(index)} function retrieves the register address from a specified index. The settings array can be saved to a binary file and subsequently loaded back into the digitizer. The \texttt{SettingExplorer} program provides a text-based interface for viewing and modifying these settings.

\subsection{\textnormal{\texttt{Data}} class}

The \texttt{Data} class is responsible for managing data from the digitizer memory (buffer), scalar information, and decoded data (energy, timestamp, waveform). It also handles data-saving operations. This class can decode data from digitizers equipped with DPP-PHA, DPP-PSD, and DPP-QDC firmware. To optimize performance, a rapid decoding function is available that extracts scalar information, energy, and timestamp data without decoding waveforms.

The decoded data is already time-sorted for each coupled-channel due to the first-in-first-out nature of the digitizer memory. To ensure limited memory usage, the decoded data is stored in a circular memory with a size of 10000 for each channel. The \texttt{Data} class requires approximately 200 MB of memory, primarily due to the circular memory. This size is sufficient to prevent data overwrites at a high trigger rate of 10 kHz/channel, assuming online data processing (like filling histograms or event building) occurs every second.

\subsection{\textnormal{\texttt{Digitizer}} class}
\label{sec::DigitizerClass}

The \texttt{Digitizer} class manages a single digitizer, storing its register settings (as described earlier) and containing an instance of the \texttt{Data} class. In a multi-digitizer system, each digitizer has its own \texttt{Digitizer} class instance. The settings array is continuously synchronized with the digitizer's actual register values and can be saved to a file for later loading. To ensure consistent settings across the digitizer, memory, and hard disk, synchronization options are available.

The \texttt{Digitizer::ReadData()} method directly retrieves the buffer from the digitizer's memory using the CAEN \texttt{CAEN\char`_DGTZ\char`_ReadData()} API and saves it to the computer's memory. The buffer is then decoded to extract energy, timestamp, and scalar information using the \texttt{Data::DecodeBuffer()} method. Additionally, the buffer can be immediately written to the hard disk when data saving is required.

Decoding the buffer in \FSUDAQ offers a significant advantage over using the CAEN \texttt{CAEN\char`_DGTZ\char`_GetDPPEvents()} API in \BoxScore. In \FSUDAQ, the decoded data is already time-sorted for each channel, unlike in \BoxScore. Additionally, energy and timestamp information are stored separately for different channels in \FSUDAQ, whereas the CAEN API combines data from different channels into a single array. This time-sorted data and channel separation simplification streamlines the histogram filling and the event-building processes.

\subsection{\textnormal{\texttt{MultiBuilder}} class}

The \texttt{MultiBuilder} class can handle event building for multiple digitizers by accessing the associated instances of the \texttt{Data} class. It contains another circular memory to store a maximum of 10000 events. Each event consists of an array of \texttt{Hit} objects. The \texttt{Hit} class holds information such as board, channel, energy, timestamp, and trace. The event-building mechanism resembles that of \BoxScore, with the distinction of allowing multiple hit occurrences for the same channel. 

A notable feature of the \texttt{MultiBuilder} class is the backward event-building capability. While event building conventionally proceeds from the earliest data, backward event building operates in the opposite direction, constructing events starting from the latest data and moving towards earlier data. This approach proves advantageous when dealing with high data rates that the normal event-building process may struggle to keep up with. The backward event building ensures that the built events always represent the latest available data, and the time consumption is controlled by limiting the number of events being constructed.

\subsection{GUI classes}

The GUI for \FSUDAQ is developed as a separate component, utilizing the core classes described earlier. Each GUI panel within \FSUDAQ is derived from the \texttt{QMainWindow} class. For example, the \texttt{FSUDAQ} class manages the main window, overseeing the overall program. The \texttt{DigiSettingsPanel} class is specifically designed for digitizer status and settings, providing a user-friendly interface for configuration. The \texttt{Scope} class enables real-time waveform inspection, while the \texttt{SingleSpectra} class facilitates online filling of decoded energy data into histograms for each channel.

To provide one- and two-dimensional histograms for displaying online data, custom histogram classes named \texttt{Histogram1D} and \texttt{Histogram2D} were developed using the \texttt{QCustomPlot} library. These classes offer essential features such as zooming, re-binning, histogram clearing, log scale, and statistical information. Additionally, the \texttt{Histogram2D} class provides graphical cut capabilities. These histogram classes enhance data visualization and online analysis, offering a user-friendly and flexible tool for exploring data distributions in both one and two dimensions.

\section{Operation of \FSUDAQ }

Once \FSUDAQ~is launched, a main thread is dedicated to handling the GUI interface. For each opened digitizer, an individual thread is created to handle data reading. Additional timing threads are also created for tasks such as updating the scalar panel, inspecting waveforms, filling single spectra, and performing online event-building when required.

Upon initiating ACQ, data reading threads are launched to continuously retrieve buffers from the digitizers. These buffers are then decoded, and if it is a data-saving run, the decoded data is immediately saved to the hard disk. The data processing rate, which includes buffer retrieval, decoding, and saving, typically ranges from 1 kHz to a few 10 kHz and is influenced by computer performance. If ACQ is started from the scope panel, rapid decoding is disabled to ensure that waveforms can be extracted into memory for plotting.

If \textsf{elog} is available, a new log entry is generated for each data-saving run, recording the start time and a user-provided start comment. The entry is subsequently updated with the stop time, stop comment, and total file size when ACQ is stopped. Regardless of \textsf{elog} availability, timing and run comments are stored in a text file named \texttt{RunTimestamp.dat} within the data directory.

The scalar panel opens automatically when ACQ starts, and a dedicated timing thread is launched to update the scalar panel every second. If an \textsf{InfluxDB} database IP address and name are provided, the trigger rates for each channel are pushed to the database.

When the scope panel is opened, waveform recording is enforced until the panel is closed, A timing thread is also initiated when ACQ started. This thread instructs the scope panel to access the most recent waveform every 200 milliseconds. Waveform plotting is a read-only operation that does not interfere with buffer retrieval or decoding. While theoretically possible to access waveforms from the \texttt{Data} class more frequently, a 200-millisecond limit is imposed to maintain a balance between performance and user experience.

\subsection{Data transfer rate and trigger rate }

The write speed of a modern mechanical hard disk is at least 100 MB/s, and that of the cable connection is 30 MB/s for USB or 80 MB/s for optical fiber, either the connection or the disk can become a hardware bottleneck for data transfer. For a single digitizer, the bottleneck is typically the cable connection. For multiple digitizers, the bottleneck is the mechanical hard disk, which can be alleviated by using a solid-state drive (SSD), as SSDs typically have write speeds exceeding 400 MB/s. 

The combined header, energy, and timestamp data only require a maximum of 36 bytes per measurement. Most of the data comes from the waveform if waveform recording is enabled through register 0x8000. Each waveform with a length of $l$ samples occupies $l \times 2$ bytes. For example, a 2000-sample waveform (equivalent to 8 $\mu$s for the V1725 digitizer) utilizes 8 kB of storage.

The data format is structured in aggregations. A readout block can contain a maximum of 1023 board aggregations, each of which can hold a maximum of 1023 coupled-channel aggregations~\cite{CAEN1730PHA,CAEN1730PSD, CAEN1740}. Each board aggregation includes a 16-byte header shared by all coupled-channels, while each coupled-channel aggregation contains an 8-byte header shared by all measurements. This structure minimizes the number of headers and increases the number of events within a single readout.

With an optical fiber connection, the maximum transfer rate is approximately 70k measurements with 496-sample waveforms or $\sim$5 million measurements without waveforms per second per fiber, or $\sim$65 MB/s/fiber (see section~\ref{Performance}.~Performance for more details).

\subsection{Time synchronization of multiple digitizers }
\label{section:Sync}

A built-in synchronization helper is provided to ensure consistent timestamps across multiple digitizers. One digitizer is designated as the master, using its internal clock, while the others rely on an external clock. To synchronize clock phases, the \textsf{CLK-OUT} of one digitizer is connected to the \textsf{CLK-IN} of the next in a daisy chain configuration, with appropriate Phase-Locked Loop (PLL) firmware installed. The synchronization helper offers four methods for clock zeroing once ACQ has started:

\begin{itemize}
    \item A) \textsf{TRG-OUT} to \textsf{S-IN} in a daisy chain,
    \item B) \textsf{TRG-OUT} to \textsf{TRG-IN} in a daisy chain,
    \item C) External trigger to the \textsf{S-IN} of the clock-reference digitizer, with the rest using \textsf{TRG-OUT} to \textsf{S-IN} in a daisy chain,
    \item D) External trigger to a fan-in/fan-out and to the \textsf{S-IN} of all digitizers.
\end{itemize}

In the \textsf{TRG-OUT} to \textsf{S-IN} (or \textsf{TRG-OUT} to \textsf{TRG-IN}) configuration, when ACQ is initiated, the last digitizer in the chain is armed first and waits for the \textsf{S-IN} (or \textsf{TRG-IN}) signal to zero its clock and begin acquisition. The first (master) digitizer, which is armed last, zeros its clock before starting the acquisition. Once the master begins ACQ, it sends a \textsf{RUN} signal through \textsf{TRG-OUT} to the \textsf{S-IN} (or \textsf{TRG-IN}) of the next digitizer, which starts its ACQ and passes the \textsf{RUN} signal to the subsequent digitizer, and so on. The advantage of this method is that it does not require an external trigger for clock zeroing.

In addition to these pre-defined synchronization options, users can customize synchronization settings using the \textsf{ACQ/Readout} and \textsf{Global/TRG-OUT/TRG-IN} tabs in the panel of digitizer settings.

While clock zeroing for each digitizer may not be perfectly synchronized due to hardware wiring and processing delays, these delays (typically a few clock ticks for setups with a few digitizers) do not significantly impact event building, which generally operates on scales of 100 ticks or more. If methods A, B, or C result in large delays due to the daisy chain configuration, method D can be employed as an alternative. Additionally, the \textsf{Run Delay} setting can be used to compensate for delays caused by the daisy chain.

\subsection{Trigger management}

\begin{figure}[ht!]
\centering
\includegraphics[trim=1cm 5.5cm 31cm 16cm, clip, width=14cm]{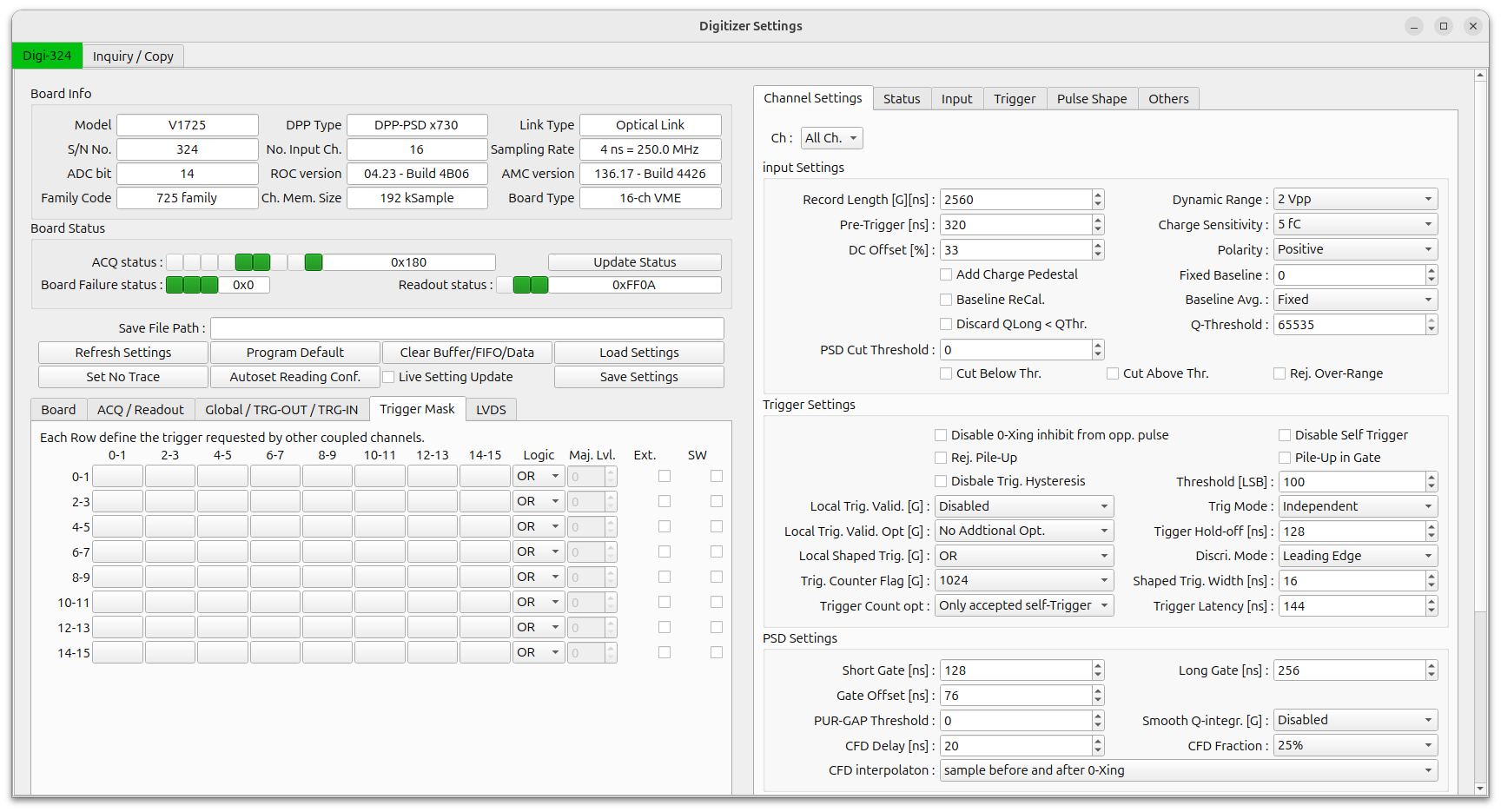}
\caption{\label{fig:TRG} A screenshot of the \texttt{Trigger Mask} tab from V1720 digitizer. Each coupled-channel can be trigger by itself, other coupled-channels, external, or software trigger. The logic can be OR, AND, or majority. Majority define the minimum number of trigger source to be validate the trigger.}
\end{figure}

CAEN digitizer offers a complex triggering system~\cite{CAEN_Trigger}. Each group coupled-channel can be triggered by any channel in that group or only triggered by minimum number of channels in that group. The coupled-channel generates a trigger request and combined with other trigger requests from other group of coupled-channel, software trigger, and \textsf{TRG-IN}, to form the Individual Trigger Logic, Global Trigger Logic, or \textsf{TRG-OUT}. The \textsf{TRG-OUT} can be used to trigger other digitizers. \FSUDAQ~ provides a graphical interface to simplify the trigger management. For example, the \texttt{Trigger Mask} tab in the panel of digitizers settings provide settings for internal trigger (Fig.~\ref{fig:TRG}).

\subsection{Memory management}

Each digitizer has a finite amount of onboard memory for storing recorded data, with each group of coupled channels sharing a single memory chip. This memory can be divided into $2^{N_b}$ parts, where $N_b$ is controlled by the aggregate organization register. Within each division, the number of events stored is determined by the number of events per aggregate, $N_e$. Only when $N_e$ events are accumulated within a division can that division be read out, allowing the next division to be filled. Once a division is read out, it is cleared.

Memory overflow may occur when the total size of $N_e$ events exceeds the capacity of a memory division. This issue is particularly significant with waveform recording, such as in the V1740D digitizer, which accommodates only 192 kSamples per channel. One method to mitigate memory overflow is to reduce the number of memory divisions, but this can potentially hinder data throughput. For example, if no divisions are used, the memory can only accommodate $N_e < 1024$ events, and a readout is only possible once $N_e$ events have been recorded.

In scenarios with a low data rate and large $N_e$, readouts may only become available every few seconds. Conversely, in high data rate situations with a small $N_e$, new data cannot be stored until memory is cleared through readouts, leading to dead time if the readout is too slow. The optimal configuration requires balancing $N_e$, $N_b$, event size, input data rate, and readout speed. A useful guideline is to ensure that $N_e$ scales proportionally with the data rate, while using at least two memory divisions to maximize memory utilization.

The \texttt{Digitizer} class includes a special method for the DPP-QDC firmware to prevent memory overflow in the x740 digitizer. This method recalculates the optimal $N_b$ before starting data acquisition, ensuring that each division has enough memory to hold $N_e$ events and that the total memory capacity is efficiently utilized.

\subsection{Online Histograms}
\label{section:OnlineHist}
Online histograms for each channel are available. There is no event building; instead, existing data are filled every second. For experiments involving many digitizers (particularly the V1740 digitizer, which has 64 channels) and high data rates, filling the histograms sequentially can be computationally demanding, potentially preventing some histograms from being updated within the one-second interval. To address this, we use a random generator to select channels and fill the corresponding histograms. The time allocated for each digitizer is also limited to ensure that all digitizers receive an equal share. Although this is not a perfect solution, it ensures that every channel has an opportunity to be filled.

\section{Integrating user analysis routine for online analysis}

The \texttt{MultiBuilder} feature in \FSUDAQ~enables online event building and allows for the customization of online analysis, including the selection of graphical gates. At the heart of this functionality is the \texttt{Analyzer} class, which incorporates essential components such as \texttt{MultiBuilder} for event building, \texttt{QMainWindow} for the graphical user interface, \texttt{QGridLayout} for managing plot and widget layouts, and a timing thread to handle both event building and plotting. Users can develop their own online analyzer by deriving a new class from the \texttt{Analyzer} class.

To customize their online analyzer, users must implement two virtual methods: \texttt{Analyzer::SetUpCanvas()} and \texttt{Analyzer::UpdateHistograms()}. The \texttt{SetUpCanvas()} method is used to define the widgets that will appear in the GUI. These widgets can include various \texttt{QWidget} elements such as \texttt{QLabel}, \texttt{QLineEdit}, \texttt{QComboBox}, \texttt{QSpinBox}, \texttt{Histogram1D}, and \texttt{Histogram2D}. This setup allows users to create a highly tailored GUI that meets their specific analysis needs. For example, combo boxes provide data selection and interfaces for specifying experimental parameters. The \texttt{UpdateHistograms()} method is where the core functions of event building, analysis, graphical gating, and histograms filling are implemented. The technical detail for implementation can be found in Ref.~\cite{FSUDAQ}.

\subsection{Coincident Analyzer}

\begin{figure}[h!]
\centering
\includegraphics[trim=0.7cm 0.6cm 0.6cm 1.9cm, clip, width=10cm]{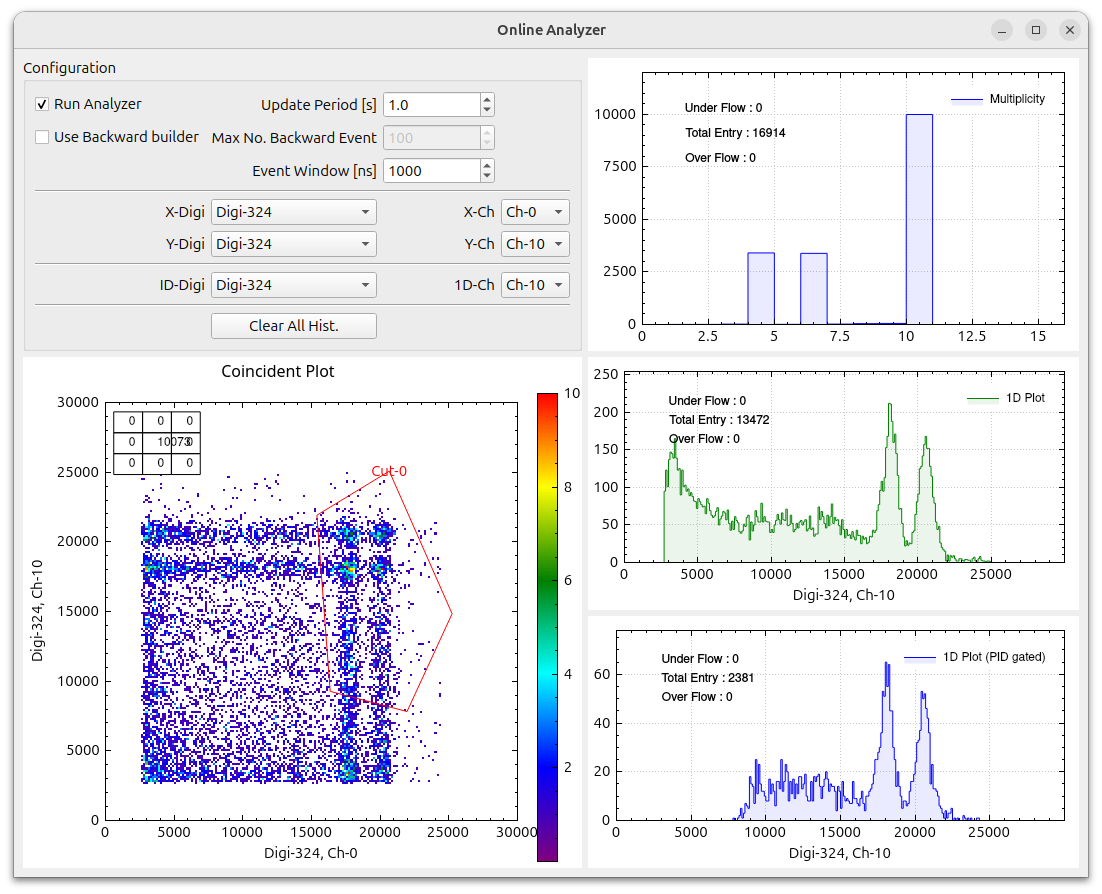}
\caption{\label{fig:CoinAna} A screenshot of the coincident analyzer panel. The control is at the upper top, the lower left is the 2D histogram for coincidences between user-selected channels. The upper left plot is a multiplicity plot, showing the statistics of the multiplicity for the events. The middle left plot is the 1D histogram for a user-selected channel. The lower left plot is the gated histogram for the 1D histogram. }
\end{figure}

A universal coincidence analyzer, derived from the \texttt{Analyzer} class, was developed to visualize coincidences between two channels. The interface, illustrated in Fig.~\ref{fig:CoinAna}, allows users to control various settings. They can specify the histogram update period and choose between two event-building methods: normal event building or backward event building. Additionally, users can select the digitizer and channel for the x- and y-axes of a coincidence 2D histogram, and a third digitizer and channel for a 1D histogram. Graphical gates can be created interactively on the 2D histogram, and once applied, the gate will filter data from the third channel accordingly.

\subsection{Neutron-Gamma Analyzer}

\begin{figure}[h!]
\centering
\includegraphics[trim=0.6cm 0.6cm 0.75cm 1.8cm, clip, width=10cm]{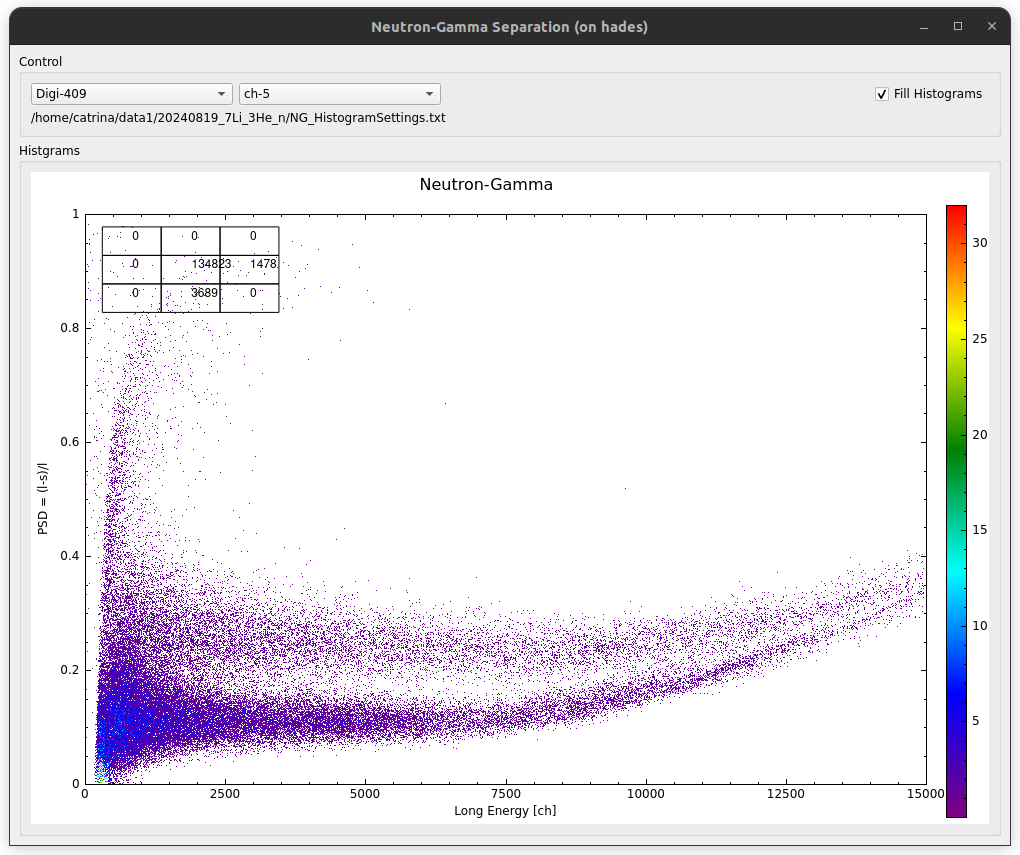}
\caption{\label{fig:NGAna} A screenshot of neutron-gamma analyzer during a testing of a neutron detector~\cite{CATRiNA} with the \iso{6}{Li}(\iso{3}{He},d) reaction. The y-axis is $(E_l-E_s)/E_l$ and the x-axis is $E_l$. The gamma is the lower band, and neutron is the upper band.}
\end{figure}

Another generic analyzer for neutron-gamma separation is available for digitizers equipped with DPP-PSD firmware. In the PSD firmware, the input signal is integrated using both a short time gate and a long time gate. Neutron-gamma separation leverages the difference in pulse shape between neutrons and gamma rays: neutrons exhibit a longer decay time, while gamma rays have a shorter decay time. The neutron-gamma analyzer is derived from the \texttt{Analyzer} class but does not employ event building. A screenshot of the online experimental data can be seen in Fig.~\ref{fig:NGAna}. This analyzer calculates the y-axis using the formula \((E_l - E_s)/E_l\), where \(E_i\) represents the energy, with the subscripts \(l\) and \(s\) denoting the long and short gates, respectively. Event building is not required for this analyzer.

\section{Testing and Deployment of \FSUDAQ}

Four nuclear physics applications of \FSUDAQ~are presented in this section. \FSUDAQ~has been tested and deployed at the John D. Fox Laboratory at FSU, as well as at the ATLAS facility at Argonne National Laboratory. The first test of \FSUDAQ~was conducted with Encore\cite{Encore}, followed by deployment on ANASEN~\cite{ANASEN}. These tests and deployments provided valuable feedback, leading to significant improvements and refinement of the system. Subsequent deployments included the Super-Enge Split-Pole Spectrograph (SPS)\cite{SPS} and the ATLAS in-flight separator system\cite{RAISOR}. In addition to these deployments, \FSUDAQ~is undergoing testing in various other setups, such as the COHERENCE collaboration at the University of North Carolina and the CATRiNA neutron array\cite{CATRiNA} at FSU.

\subsection{\textnormal{Encore} active-target detector at FSU}

The \FSUDAQ~was first deployed with the Encore active-target detector~\cite{Encore}, which operated at a pressure of 50 to 70 Torr with a composite gas mixture containing carbon, oxygen, nitrogen, and hydrogen in a weight ratio of 0.46:0.41:0.10:0.04. A \iso{12}{C} beam of approximately 3000 pps, with energy ranging from 5 to 20 MeV/u, was directed at the detector. The setup included 34 channels, each with an average trigger rate of 2 kHz. The signals were first sent to a preamplifier and then delivered to three digitizers running the DPP-PHA firmware, all of which used self-triggering. Only energy and timestamp data were recorded. Each digitizer was connected to the DAQ computer via a dedicated optical cable. The DAQ computer, located near Encore, was linked to the laboratory's Ethernet network with a 1 Gb/s connection, while the \textsf{elog} and \textsf{InfluxDB} database were hosted on a separate computer within the network.  The \textsf{elog} and \textsf{InfluxDB} database were hosted on a separate computer within the network, and the DAQ system was operated remotely via an SSH connection from the control room. Time synchronization between the digitizers was achieved using the \textsf{TRG-OUT} to \textsf{S-IN} method described earlier.

\begin{figure}[ht!]
\centering
\includegraphics[trim=1.0cm 3.5cm 1.0cm 9.0cm, clip, width=10cm]{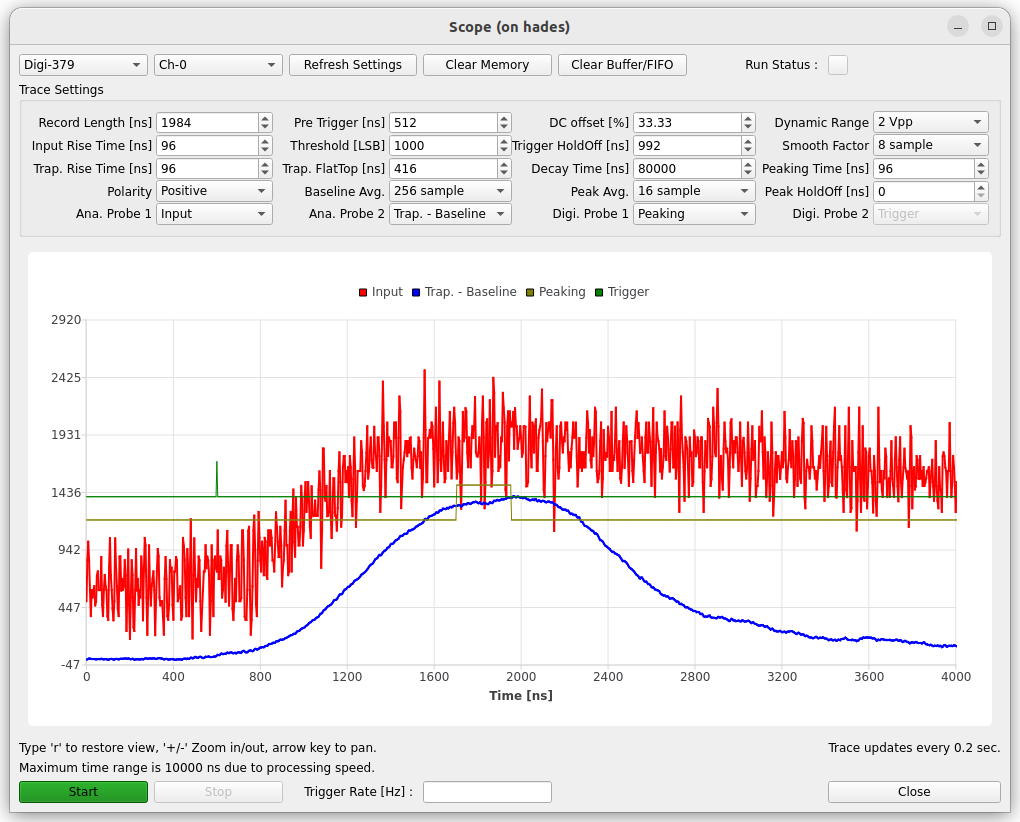}
\caption{\label{fig:EncoreTrace} A screenshot of the scope panel during the experiment. The signal-to-noise ratio is roughly 3:1. The red line is the input signal, the blue line is the baseline subtracted trapezoid retrieved from the digitizer. The brown line indicates the integration range from the trapezoid. The green line indicates the trigger time. 
}
\end{figure}

During the experiment, a 5 Hz pulser signal was fed into the preamps to monitor synchronization and energy gain stability throughout the run. No detectable timing offsets were observed between the digitizers using the \textsf{TRG-OUT} to \textsf{S-IN} synchronization method. The total data rate was approximately 3 MB/s, with only energy and timestamps recorded. The signal had a long rise time of 1 µs, and the signal-to-noise ratio varied from 1:1 to 5:1. A sample trace from the experiment is shown in Figure~\ref{fig:EncoreTrace}.

\begin{figure}[ht!]
\centering
\includegraphics[trim=0.5cm 28cm 0.5cm 2.5cm, clip, width=10cm]{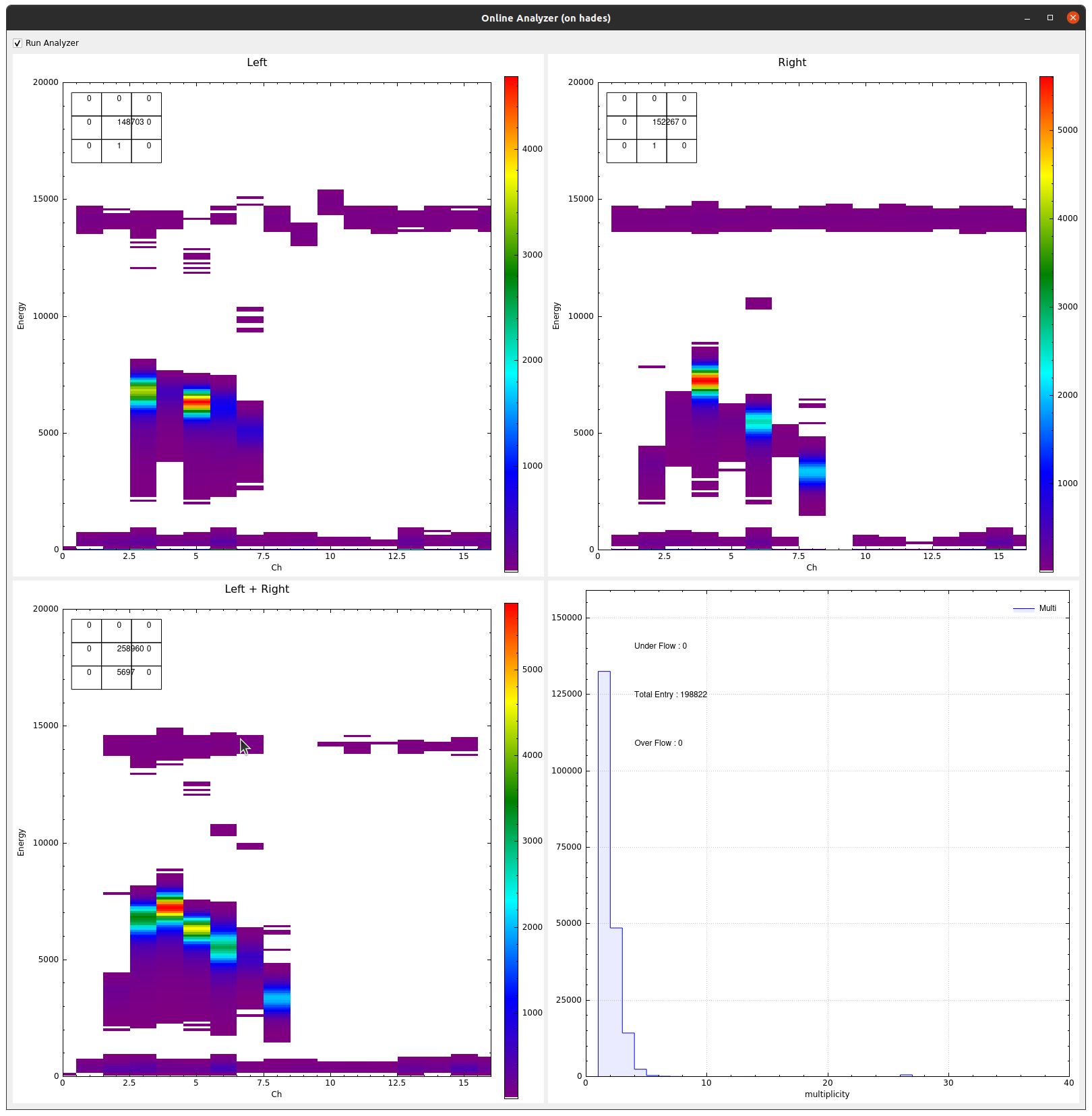}
\caption{\label{fig:EncoreOnline} A screenshot of the online analyzer of Encore during the experiment. The beam came from left to right. The Bragg peak was at the 3rd strip. The horizontal band on the top is the signals from the 5 Hz pulser.}
\end{figure}

An online analyzer, \texttt{EncoreAnalyzer}, was developed by inheriting from the \texttt{Analyzer} class mentioned earlier. The event-building window was set to 8 µs. The event-building and histogram filling event one second. The analyzer included mapping and energy gain-matching parameters, which were deduced through semi-online analysis (discussed below). Apart from the gain-matching, no additional calculations were performed. Four histograms were generated and plotted: the left strips, right strips, the sum of both, and the multiplicity. The online analyzer operated smoothly at approximately 2 kHz across all channels, with a typical canvas shown in Figure~\ref{fig:EncoreOnline}.

In addition to the online analysis, a semi-online analysis was employed to process data files using an offline event builder, which come with the \FSUDAQ. This event builder converted raw data into ROOT files, which were then analyzed, and histograms were plotted, allowing the deduction of gain-matching parameters.

In summary, although the testing process exposed various issues, numerous fixes were successfully applied during the experiment. The \FSUDAQ~tested with Encore performed effectively, achieving synchronized timing across digitizers and problem-free data acquisition. Its online analyzer (or semi-online analyzer) provided essential real-time information, enabling experimenters to monitor the experiment's status and make informed decisions.

\subsection{\textnormal{ANASEN} at FSU}

A second deployment of \FSUDAQ was performed on the ANASEN active target~\cite{ANASEN}, equipped with tracking capabilities. This configuration utilized eleven digitizers, employing a mix of firmware and models. Seven V1740D digitizers (64 channels each) with DPP-QDC firmware were connected to position-sensitive silicon detectors. Any triggered channel generated a \textsf{TRG-OUT} signal, which initiated data acquisition in three V1725 digitizers with DPP-PHA firmware (not all channels were used) and one V1730 digitizer with DPP-PSD firmware. The acquisition start signal armed the digitizers, while time synchronization was achieved using a gate generator connected to a logical Fan-In Fan-Out unit, propagating the zeroing signal to all eleven digitizers (method D in section~\ref{section:Sync}). Four optical cables were used to connect the digitizers to the computer, with the seven V1740D digitizers grouped into three daisy-chained configurations, sharing a single cable each. The remaining four digitizers were also connected in a daisy chain. A total of 496 channels were in use. ANASEN recorded only energy and timestamps, without waveform traces. Although no dedicated online analyzer was implemented for this experiment, online single histograms, directly filled from the data without event building, were used to monitor data collection.

Two experiments, the $^{29}$Mg($\alpha$,p) and $^{26}$Al($d,p$) reactions, were conducted using \FSUDAQ. Data analysis is currently in progress. Preliminary analysis suggests that the data acquisition and time synchronization between the digitizers were functioning correctly. During the experiments, trigger rates varied substantially across channels, ranging from a few Hertz to several tens of kilohertz.

The ANASEN deployment provided valuable insights, leading to several enhancements in \FSUDAQ. One notable improvement was the implementation of random filling for online single histograms (see section~\ref{section:OnlineHist}).

\subsection{Super-Enge Split-Pole Spectrograph at FSU}

Following successful testing in Encore and ANASEN, the core functionality of \FSUDAQ~is well-established, including smooth data acquisition, robust digitizer settings, and effective online scope and single histograms. The final component to be tested is a slightly more complex online analyzer.

The $^{12}$C($d,p$) reaction at 8 MeV/u was conducted using the Split-Pole Spectrograph~\cite{SPS} in conjunction with \FSUDAQ. A pure deuteron beam with an intensity of 10 nA ($6.2\times10^{10}$ pps) bombarded a 50 $\mu$g/cm$^2$ thick enriched $^{12}$C target. The spectrograph was set to 20 degrees with a field of 0.75 T. The trigger rates for the plastic scintillator and focal plane detector were approximately 3 kHz and 2 kHz, respectively.

The setup consisted of only 9 channels: two channels for the left and right PMTs of the plastic scintillator, and seven channels for the focal plane detector. A single VX1730(S) digitizer with DPP-PSD firmware was used for data acquisition, and all channels were self-trigger.

\begin{figure}[ht]
\centering
\includegraphics[trim=0.5cm 0.5cm 0.5cm 13.4cm, clip, width=10cm]{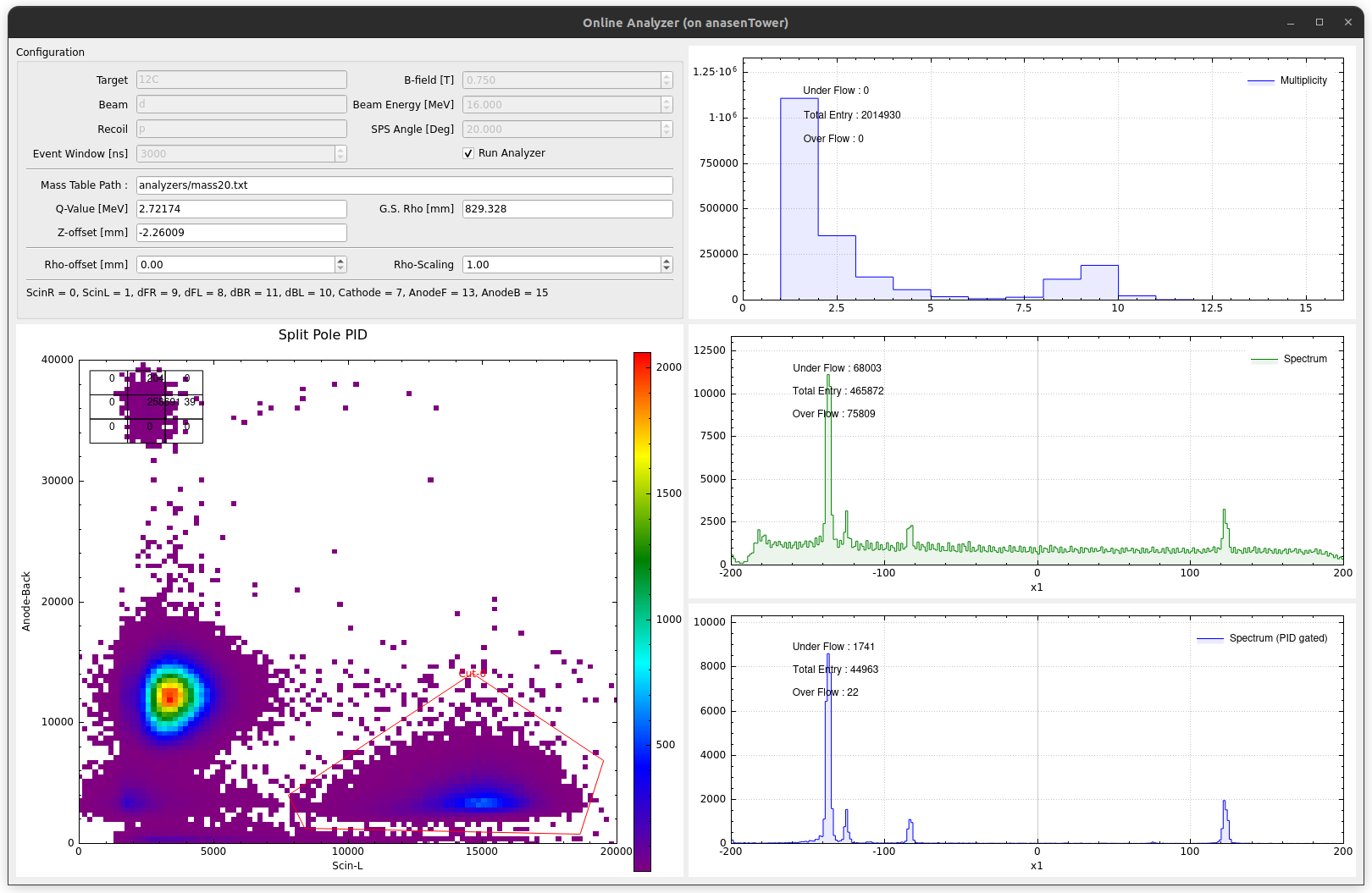}
\caption{\label{fig:splitpole} A screenshot of the online analyzer of the SPS. There is a red polygon, which was drawn online, for the proton gate on the left plot. see text for detail. The \iso{13}{C} ground state (at $\sim$120) and first three excited state can be seen on the right plots.}
\end{figure}

Similar to the Encore application, a \texttt{SplitPoleAnalyser} class, inheriting from the \texttt{Analysis} class, was developed for online analysis. A standard event building process with a 3 $\mu$s event window was employed. The online analyzer's canvas is depicted in Figure~\ref{fig:splitpole}. The left panel shows the particle-identification (PID) plot using the energy loss of the focal plane back-side anode as $\Delta E$ and the energy loss of the left PMT of the scintillator for $E$. The upper right panel displays the $x_1$ position, while the lower right panel shows the $x_1$ position after applying the PID gate, drawn online on the PID plot. The online event builder calculates the focal plane spectrum and also apply a graphical cut.

The test was successful, demonstrating robust data acquisition and the online event builder's ability to construct the PID plot and apply an online graphical gate.

\subsection{ATLAS in-flight system at ANL}

The \BoxScore system, currently used for online beam tuning and diagnosis in the ATLAS in-flight system~\cite{RAISOR}, faces limitations related to its online event builder and reliance on CERN ROOT. These factors can impact performance, especially at high data rates. Furthermore, \BoxScore~is restricted to a single digitizer.
\FSUDAQ~aims to overcome these limitations by providing a more versatile framework that supports various beam diagnosis devices, including microchannel plates (MCPs).

\begin{figure}[ht]
\centering
\includegraphics[trim=0.cm 0.cm 0.cm 0.cm, clip, width=10cm]{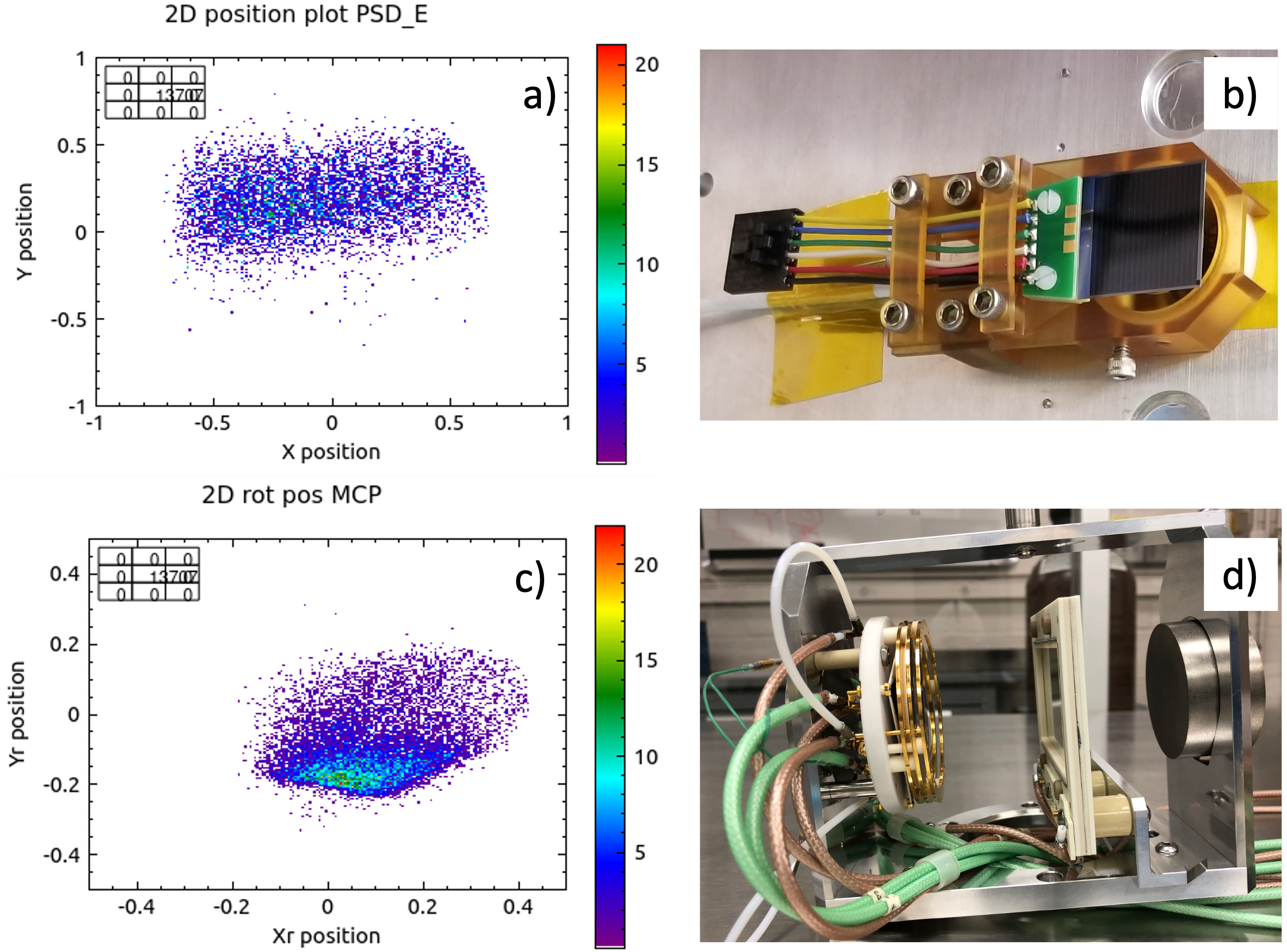}
\caption{\label{fig:MCP_PSD} a) A screenshot of the online beam profile from the b) position sensitive silicon detector. c) A screenshot of the online beam profile from the c) MCP.}
\end{figure}

The built-in coincident analyzer can be utilized as a PID plot for radioactive beam tuning. It can be modified to output PID gates to an external database and share the rates of isotopes. For beam profile measurements, a new analyzer class, \texttt{RaisorAnalyzer}, was created. This class can process multiple channels and calculate the x-y position from a position-sensitive PSD~\cite{MSPSD} or MCP~\cite{MCP}. Figure~\ref{fig:MCP_PSD} shows a screenshot of the online beam profile display and the photos of the detectors. The deployment and implementation of \FSUDAQ~is currently under detail testing.

\section{Performance} \label{Performance}

A single V1725 digitizer with DPP-PHA firmware was utilized for these tests. \FSUDAQ's performance was evaluated on a \mbox{Ubuntu 22.04} machine equipped with an Intel\textsuperscript{\textregistered}~Core\texttrademark~i7-7700 CPU (3.6 GHz, 8 cores), 64 GB RAM, and a 1 TB NVMe SSD hard-disk.

Without any digitizer connected, \FSUDAQ~consumed approximately 90~MB of memory. When a digitizer was connected, an additional \mbox{200 MB} of memory was used.

Input signals were generated using a CAEN DT5810B signal emulator. Each pulse had a rise time of 10 ns, a decay time of 500 ns, and an amplitude of 1 volt. The pulses were split into 10 channels using a linear fan-in-fan-out (CAEN V925), and each channel was connected to the digitizer.

The digitizer's sample size was set to 496 samples (1984 ns), with an input rise time of 96 ns, trigger and peak hold-offs of 0 ns, and a threshold of 1000 least-significant-bits. The trapezoid settings were 96 ns for the rise time, 96 ns for the flat-top, 500 ns for the pole-zero, and 0 ns for the peaking time. Under these conditions, the theoretical maximum pile-up free trigger rate is 3.47 MHz. Database output was enabled, and the number of events per aggregation was set to 1023 (register 0x1034), the maximum number of aggregations per readout was 511 (register 0xEF1C), and the aggregation organization was 511 (register 0x800C) to maximize data readout. 

The scope functioned as expected when only one channel was enabled at a time, and the online histograms operated without issues. However, the \texttt{MultiBuilder} component had not been tested at this high data rate.

\begin{figure}
    \centering
    \includegraphics[trim=0cm 0cm 0cm 0.1cm, clip, width=7cm]{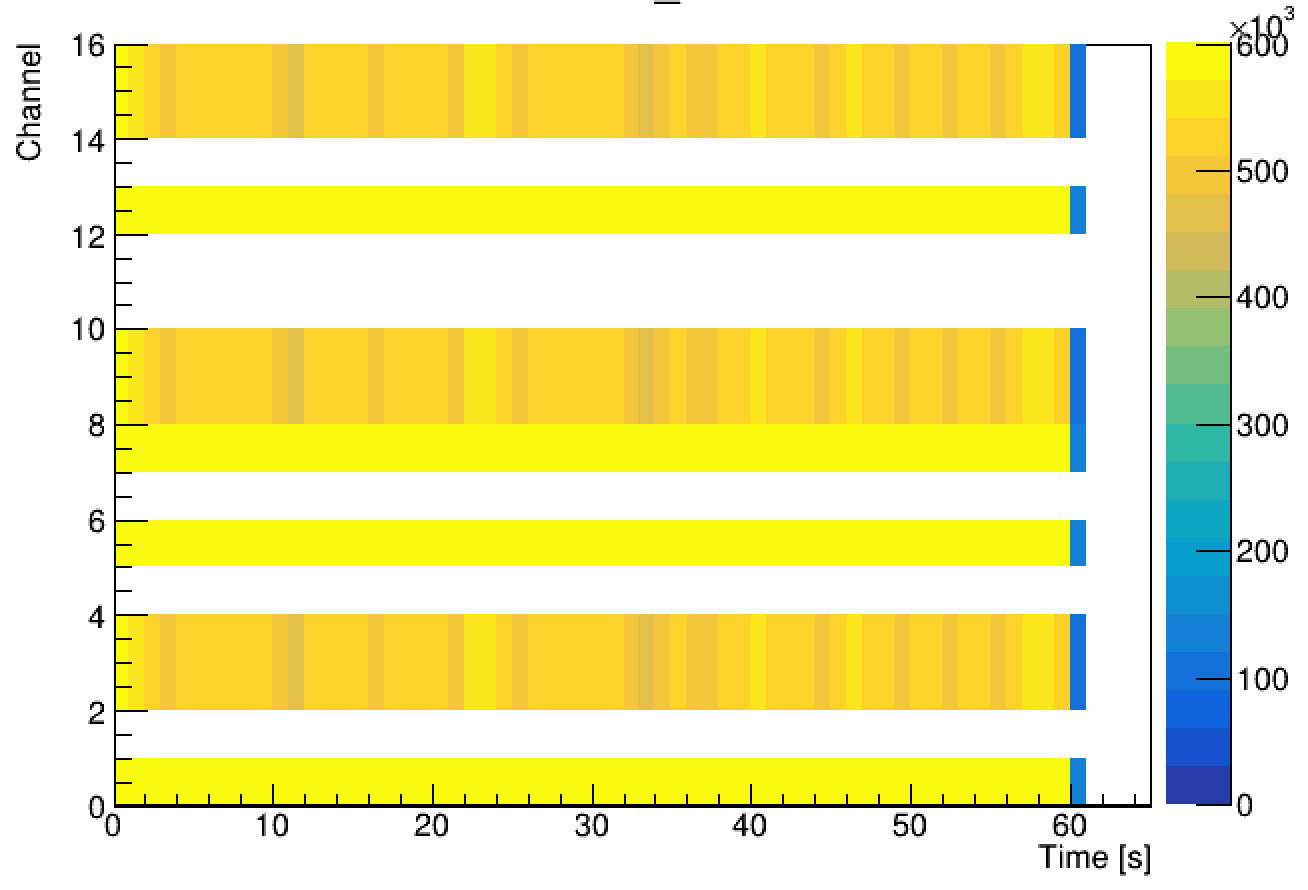}
    \caption{The trigger rate of 600 kHz input to 10 channels of V1725 digitizer. See text for more detail.}
    \label{fig:triggerRate}
\end{figure}

Without waveform saving, a 600 kHz signal was sent to 10 channels. The data transfer rate was $\sim$65 MB/s. After converting the data into a ROOT file, the trigger rate for all input channels is shown in Figure~\ref{fig:triggerRate}. The onboard memory of the digitizer is shared between channels 2i and 2(i+1). Channels 0, 5, and 12 have exclusive memory access, allowing them to handle 600 kHz, while the other channels, due to memory sharing, experience reduced performance to approximately 500 kHz/channel. With waveform saving of 496 samples, the digitizer was able to accept 70 kHz/channel. The data transfer rate remained limited to 65 MB/s.

\section{Including other CAEN digitizers with different DPP firmware}

The DPP-PHA firmware was initially developed for \FSUDAQ. Subsequently, DPP-PSD and DPP-QDC firmware support were added. Implementing DPP-PSD support required four key modifications:

\begin{itemize}
\item Adding the register addresses to the \texttt{RegisterAddress.h} file in the \texttt{Reg} class.
\item Implementing the data decoding method for DPP-PSD firmware in the \texttt{Data} class.
\item Adding the GUI interface for DPP-PSD settings in the \texttt{DigSettingPanel} class and the \texttt{Scope} class.
\end{itemize}

Similar modifications were applied to include DPP-QDC support. This modular approach suggests that integrating other DPP firmwares and digitizer models should be feasible in the future. However, the V1743 digitizer, which does not utilize DPP firmware and is controlled via the CAEN Digitizer API, presents a unique challenge due to the lack of an open register table for direct control. In such cases, one approach would be to create a derivative class of the \texttt{Digitizer} class that leverages the CAEN Digitizer API to control and read out the digitizer.

\section{Summary}

\FSUDAQ~was designed to be a versatile and user-friendly data acquisition program for various nuclear physics experiments. It is specifically tailored for the CAEN V1725, V1730, and V1740 digitizers equipped with DPP-PHA, DPP-PSD, and DPP-QDC firmware.
\FSUDAQ offers a graphical interface for controlling digitizers by manipulating their onboard register values. Users can view online waveforms through the scope panel and adjust related settings. Additionally, online single spectra for each channel are available, and online event builders can be constructed and customized. The program can also connect to an external \textsf{InfluxDB} database and \textsf{elog}.
\FSUDAQ is capable of recording data at approximately 500 kHz/channel without waveforms and 70 kHz/channel with 495 sample waveforms. It can also be run on Raspberry Pi 4B or 5 using USB or A4818 optical-to-USB adapters. For more technical details, please refer to the wiki page~\cite{FSUDAQ_WIKI}. Additionally, a Discord channel~\cite{Discord} is available for support.

\section{Acknowledgement}
The author would like to acknowledge the support and operations staff at FSU (E. Lopez-Saavedra and S. Almaraz-Calderon for the Encore, M. C. Spieker, B. Kelly, and A. Conley for the SPS, I. L. Wiedenhoever and V. Sitaraman for the ANASEN, L. baby, P. Barber, and B.~Schmidt for delivering beams) and ANL (K. J. Bhatt, I. Tolstukhin, and C. R. Calem for RAISOR). This research used resources from Florida State University's John D. Fox Laboratory, which is supported by National Science Foundation, and Argonne National Laboratory’s ATLAS facility, which is a Department of Energy Office of Science User Facility. This work was supported by the National Science Foundation under Grant No. PHY-2012522. This material is based upon work supported by the U.S. Department of Energy, Office of Science, Office of Nuclear Physics, under Contract No. DE-AC02-06CH11357.



\end{document}